\documentclass[showpacs,amssymb,floatfix,twocolumn]{revtex4}
\usepackage{graphicx}
\usepackage{dcolumn}
\usepackage{bm}
\usepackage{mathrsfs}
\begin{document}
\title{d-Wave Pairing in an Ensemble of Spin Polaron
Quasiparticles in the Spin-Fermion Model of the Electronic
Structure of the CuO$_2$ Plane}

\author{V.\,V.\,Val'kov$^{a,b)}$,
        D.\,M.\,Dzebisashvili$^{a,b)}$,
        A.\,F.\,Barabanov$^{c)}$}

\address{$^{a}$ L. V. Kirensky Institute of Physics SB RAS, 660036 Krasnoyarsk, Russia \\
         $^{b}$ Siberian State Aerospace University, 660014 Krasnoyarsk, Russia\\
         $^{c}$ Institute for High Pressure Physics, 142098 Troitsk, Moscow
         Region, Russia\\
         vvv@iph.krasn.ru, ddm@iph.krasn.ru, abarab@bk.ru}
\date{\today}
\begin{abstract}
It is demonstrated for the first time that the strong coupling
between spin moments of copper ions and oxygen holes, which arises
upon hybridazation mixing of two hole subsystems in the Emery
model, not only affects the formation of spin polaron
quasiparticles but also ensures effective attraction between them
via the exchange interaction. This results in the Cooper
instability with d-wave pairing in a 2D ensemble of spin polaron
quasiparticles. The $T-x$-phase diagram obtained using this
approach agrees well with the available experimental data.
\end{abstract}
 \maketitle

1. \textbf{Introduction}

Study of high-temperature superconductors (HTS) revealed the
importance of the interactions between charge and spin degrees of
freedom \cite{Plakida_book_10}. These interactions manifest
themselves in the properties of both the normal HTS phase,
leading, e. g., to the pseudogap behavior \cite{Sadovskii01}, and
the superconducting state, ensuring, in particular, d-type
order-parameter symmetry \cite{Izumov97}.

Many theoretical works on the spin-fermion interactions were based
on the Hubbard model \cite{Zaitsev_2004,Plakida03,Vladimirov07}
and $t-J$- and $t-J^*$- models
\cite{Kagan_1994,VVDO_2002,Valkov2005,Ovchinnikov_2009} in which
both the charge and spin subsystems were formed by the same
electrons.

Meanwhile, in the real structure of the HTS's CuO$_2$-plane, the
spin moments of copper ions and holes that move over oxygen ions
are spatially separated. In addition, the presence of two oxygen
ions in the unit cell causes a number of features.

As is known, the multi-band character of the Emery model
\cite{Emery1987,Varma_1987,Hirsch_1987} makes the theoretical
consideration of the Copper instability extremely intricate.
Therefore, the conditions for implementation of the
superconducting d-phase with regard to the above-mentioned
features of the CuO$_2$-plane structure require further
investigations.

The aim of this study was to develop the theory of the 2D
superconducting phase with d-type order-parameter symmetry that
would be free of the above-said limitations. It is important that
the conditions for implementation of the Copper instability will
be determined for an ensemble of the spin polaron quasiparticles
formed by the strong coupling between the spin moments of copper
ions and oxygen holes.
\bigskip

2. \textbf{Hamiltonian of the spin-fermion model of the CuO$_2$-
plane}

As is known, smallness of parameter $t_{pd}$ of mixing between the
p-states of oxygen ions and d-states of copper ions as compared
with the difference $\Delta_{pd}=\varepsilon_p-\varepsilon_d$
between the energies of these states in the strong correlation
regime allows to obtain the SU(2)-invariant spin-fermion model
\cite{Zaanen_1988,Barabanov_PRB1997}:
\begin{eqnarray}\label{Heff}
\hat{\mathscr H}=\varepsilon_p\sum_lc^+_lc_l
-t\sum_{l\rho}c^+_lc_{l+\rho}+~~~~\nonumber\\
+ \sum_{f\delta \delta ^{\prime }}c_{f+\delta }^{+}\left[
\frac{\tau_-}{2} +\tilde{S}_{f}\tau_+\right] c_{f+\delta ^{\prime
}} +\hat{\mathscr H}_{exch},
\end{eqnarray}
where
$$\tau_\pm=\tau(1\pm\eta),~~\tau =\frac{(t_{pd})^{2}}{\Delta
_{pd}},~~\eta=\frac{\Delta_{pd}}{U_d-\Delta_{pd}},~\tilde{S}_{f}=\vec
S_{f}\vec\sigma.$$

The first term in Eq. (\ref{Heff}) describes the energy of
coupling between a doped hole and an oxygen ion. Hereinafter,
energy $\varepsilon_p$ is assumed to be counted from chemical
potential $\mu$. The operator $c^{+}_{l}=(c^{+}_{l\uparrow
},c^{+}_{l\downarrow})$ in the spinor representation describes
creation of a hole on an oxygen ion with site number $l$.

The second term in $\hat {\mathscr H}$ corresponds to direct
hoppings of holes between the nearest oxygen ions coupled by
vectors $\rho$. The intensity of these hoppings is determined by
the tunneling integral $t>0$. Hereinafter, hybridization parameter
$t_{pd}$ is assumed to exceed the tunneling integral, $t_{pd}>t$.

The third term in (\ref{Heff}) is caused by the account for the
second-order processes by hybridization parameter $t_{pd}$. The
arising operator describes hoppings of a hole between oxygen ions
adjacent to the copper ion. Operator $\tilde{S}_{f}$ is determined
as a scalar product of vector operator $\vec S_{f}$ of the spin
moment on the copper ion in the site with index $f$ and the vector
$\vec \sigma =(\sigma ^{x},\sigma ^{y},\sigma ^{z})$ that consists
of the Pauli matrices. The most important feature of this operator
is that it contains contributions of hoppings of holes accompanied
by the spin-flip processes. At such hoppings, spin projections of
both the hole and the copper ion change. The account for these
contributions strongly affects the structure of the spin polaron
elementary excitation spectrum. Vectors $\mathbf{\delta }$ and
$\mathbf{\delta }^{\prime }$ independently acquire the four values
$\{\pm a_{x},\pm a_{y}\}=\frac{1}{2}\{\pm g_{x},\pm g_{y}\}$,
where $\{\pm g_{x},\pm g_{y}\}$ are the vectors of the nearest
neighbors of the copper lattice.

The last term in (\ref{Heff}) describes the exchange interaction
between spins of copper ions. Hereinafter, we limit the
consideration to the interactions of the spins located within two
coordination spheres:
\begin{eqnarray}\label{HJ}
\hat {\mathscr H}_{exch}=\frac{I_{1}}{2}\sum_{fg}\vec S_{f}\vec
S_{f+g}+\frac{I_{2}}{2}\sum_{fd}\vec S_{f}\vec S_{f+d}.
\end{eqnarray}
Here, $I_1$ and $I_2~$ are the exchange integrals for the nearest
and next-nearest ($d=\pm g_{x}\pm g_{y}$) spins, respectively. The
exchange constants are convenient to express via frustration
parameter $p$ and effective exchange integral $I$:
\begin{eqnarray}
I_{1}=(1-p)I,~~~I_{2}=pI,~~~0\leq p\leq 1,~~I>0.
\end{eqnarray}
Quantity $p$ can be related to concentration of holes $x$ per
copper atom \cite{Barabanov_JETP2001}.

Note that in reality the hopping integrals in the first and second
terms of Hamiltonian (\ref{Heff}) can have different signs for
different hopping directions. It is easy to demonstrate that these
signs can be taken into account by introducing the factors $\exp
\{iQ(l-{l^{\prime }\}}$, where $Q=(\pi,\pi)$. After the unitary
transformation $e^{iQl}c_{l}\rightarrow c_{l}$, these factors
vanish and the spectrum can be reconstructed by the shift
$k\rightarrow k+Q$ in the $k$-space.

Below we use the following commonly accepted parameter values:
$t_{pd}=1.3$ eV, $\Delta_{pd}=3.6$ eV, $U_d=10.5$ ýÂ eV, and
$t=0.1$ eV \cite{Ogata2008}. For these values, we have $\tau=0.47$
eV and $\eta=0.52$.
\bigskip

3. \textbf{Fermi quasiparticles in the strong coupling regime.}

To clarify the nature of the Fermi quasiparticles that arise in
the CuO$_2$-plane upon light doping, we consider the solution of
the Schrodinger equation for one hole by the variational
technique. We will take into account that, in accordance with the
Mermin--Wagner theorem \cite{Mermin_Wagner_1966}, without doping
at an arbitrarily low temperature the 2D subsystem of the
localized spin moments is in the state $|G\rangle$ without the
long-range magnetic order. At the antiferromagnetic exchange
interaction, this state is characterized by the properties
\cite{BarMikhShvar2011}:
\begin{eqnarray}
{\vec S}^2_{tot}|G\rangle=0|G\rangle,~ \langle
G|S_{f}^{x,y,z}|G\rangle=0,~ \vec S_{tot}=\sum_f\vec S_f.
\end{eqnarray}
The assumption about the singlet character of the state of the
considered 2D system at a finite temperature is based on the
results reported in \cite{Marshall_1955}, where it was
mathematically strictly demonstrated that the ground state of a
system of an arbitrarily large yet finite number of localized
spins in the sites of the square lattice that
antiferromagnetically interact with one another is singlet (the
Marshall theorem).

Taking into account the symmetrical properties of the Hamiltonian,
we obtain that for each irreducible representation $k$ of the
translations group, the state with one hole
$|\psi_{k\sigma}\rangle$ with spin moment projection $\sigma$ can
be written as:
\begin{eqnarray}\label{PSIk}
|\psi_{k\sigma}\rangle=\sum_{j}\alpha_{jk}
A^+_{jk\sigma}|G\rangle,
\end{eqnarray}
where $A^+_{jk\sigma}$ are, defined below, basis operators.

According to the stationarity condition of the energy functional
under the additional condition
$\langle\psi_{k\sigma}|\psi_{k\sigma}\rangle=1$, with the use of
the Lagrange technique, we obtain that the excitation energies
$\varepsilon_k=E_k-E_G$, where $E_k$ and $E_G$ are the energies of
the $|\psi_{k\sigma}\rangle$ and $|G\rangle$ states, respectively,
and coefficients $\alpha_{jk}$ are determined by the system of the
linear homogeneous equations:
\begin{eqnarray}\label{sysEqVar}
\sum_j\left[ D_{ij}(k)-\varepsilon_k
K_{ij}(k)\right]\alpha_{jk}=0,
\end{eqnarray}
where
\begin{eqnarray}\label{Dij}
D_{ij}(k)=\langle G|\{[ A_{ik\sigma},\hat{\mathscr
H}]_{-},A^+_{jk\sigma}\}_{+}|G\rangle,\\ \label{Kij}
K_{ij}(k)=\langle G|\{ A_{ik\sigma},A^+_{jk\sigma}\}_{+}|G\rangle.
\end{eqnarray}
The numerical calculations showed that the states of the one-hole
sector can be described optimally, in terms of reaching the lowest
energy at the minimum set of basis operators, when we limit the
consideration to the three families of operators:
\begin{eqnarray}\label{BO1_3}
A_{1(2)f\sigma}\equiv c_{f+a_{x(y)},\sigma},~ A_{3f\sigma}=\frac12
\sum_{\delta}\left(\tilde S_fc_{f+\delta}\right)_\sigma,~
\end{eqnarray}
used in building the operators in the quasimomentum representation
$$A_{jk\sigma}=N^{-1/2}\sum_{f}e^{-ikf}A_{jf\sigma},~(j=1,2,3).$$
Note that the further increase in the number of basis operators
almost does not affect the spin polaron branch of the spectrum.
The calculations yield ($K_{ij}=\delta_{ij}K_{ii}$):
\begin{eqnarray}\label{K13}
K_{11}(k)=K_{22}(k)=1,~ K_{33}(k)=\frac34+C_1\gamma_1(k),\\
\label{D13} D_{11(22)}=\varepsilon_p+\tau_-\left(1+\cos
k_{x(y)}\right),
\nonumber\\
D_{12}=D_{21}^*=\left( \frac{\tau_-}{2}-t
\right)\left(1+e^{ik_x}\right) \left(1+e^{-ik_y}\right),
\nonumber\\
D_{1(2),3}=D_{3,1(2)}^*=2\tau_+K_{33}\left(1+e^{ik_{x(y)}}\right),
\nonumber\\
D_{33}=\left(\varepsilon_p-2t+\frac52\tau_--4\tau_+\right)K_{33}+
\nonumber\\
+(\tau_--2t)\left(C_1\gamma_{1k}+C_2\gamma_{2k}\right)+ \nonumber\\
+\frac{\tau_-}{2}\left(C_1\gamma_{1k}+C_3\gamma_{3k}\right)+
\tau_+C_1(1-4\gamma_{1k})+ \nonumber\\
+I_1C_1(\gamma_{1k}-4)-4I_2C_2.
\end{eqnarray}
Here, $\gamma_{jk}$  ($j=1,2,3$) are the square lattice
invariants: $\gamma_{1k}=(\cos k_x+\cos k_y)/2$, $\gamma_{2k}=\cos
k_x \cos k_y,$ $\gamma_{3k}=(\cos 2k_x+\cos 2k_y)/2. $

Spin correlators $C_1$, $C_2$ and $C_3$ are averaged by the
$|G\rangle$ state of the SU(2)-invariant products of spin
operators corresponding to different cells: $C_{j}=\langle G| \vec
S_{f}\vec S_{f+r_j}|G\rangle $, where $r_j$ is the radius of the
$j$-th coordination sphere. The SU(2) invariance of the
$|G\rangle$ state leads to the equality:
\begin{eqnarray}\label{spcor}
 C_{j}=3\langle S_{f}^xS_{f+r_j}^x\rangle=3\langle
S_{f}^yS_{f+r_j}^y\rangle=3\langle S_{f}^zS_{f+r_j}^z\rangle.
\end{eqnarray}
As can be seen from the results of the numerical calculation (
Figs. 1 and 2), it is important to take into account the
interaction between the spin and charge degrees of freedom, as
well as the spin polaron character of the lower branch of the
one-hole state spectrum. The left part in Fig. 1 shows the
one-hole state energy spectrum obtained with the use of the only
two operators, $A_{1k\sigma}$ and $A_{2k\sigma}$, for the
quasimomenta located on the main diagonal of the Brillouin zone.
These branches describe, in fact, the spectrum of holes that do
not interact with the subsystem of the spin moments of copper
ions.

\begin{figure}[h]
\begin{center}
\includegraphics[width=260pt, height=200pt, angle=0, keepaspectratio]{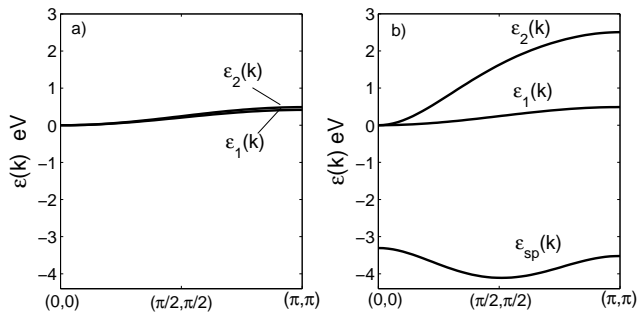}
\caption{\label{Fsp} One-hole state energies $\varepsilon_{jk}$ vs
quasimomentum along the main diagonal of the Brillouin zone for
the parameters $\tau=0.47$ eV, $\eta=0.52$, $t=0.1$ eV, and
$I=0.2\tau$. (a) Energy spectrum obtained using two operators
$A_{1k\sigma}$ and $A_{2k\sigma}$ and (b) one-hole state energies
calculated in the basis of three operators $A_{1k\sigma}$,
$A_{2k\sigma}$ and $A_{3k\sigma}$. The lower branch of the
spectrum corresponds to the spin polaron states.}
\end{center}
\end{figure}

Addition of the third operator, $A_{3k\sigma}$, to the variational
procedure leads to important qualitative transformations, which is
demonstrated by the one-hole state spectrum obtained in the basis
of three operators (Fig. \ref{Fsp} on the right). The main
difference is the arising split branch with the minimum near
$(\pi/2,\pi/2)$. The energy of such one-hole states lowers due to
the Hamiltonian term $\sim\tau_+$, which describes both the
exchange interaction between the hole and the nearest copper ions
and the spin-correlated hoppings. Inclusion in the operator basis
of the operators that explicitly take into account this strong
spin-fermion correlation yields a significant energy gain. In
addition, this is accompanied by renormalization of the two bare
branches of the spectrum.

It should be noted that, physically, the occurrence of the spin
polaron states is analogous to their occurrence in the exactly
solved problem on one electron with the spin flip in a
ferromagnetic matrix at the antiferromagnetic s-d exchange
coupling between the electron spin and the localized spin moment
\cite{Izyumov_1970}.

The above-mentioned qualitative modifications in the energy
spectrum are kept at the quasimomentum variation in other
directions of the Brillouin zone.
\begin{figure}[h]
\begin{center}
\includegraphics[width=230pt, height=200pt, angle=0, keepaspectratio]{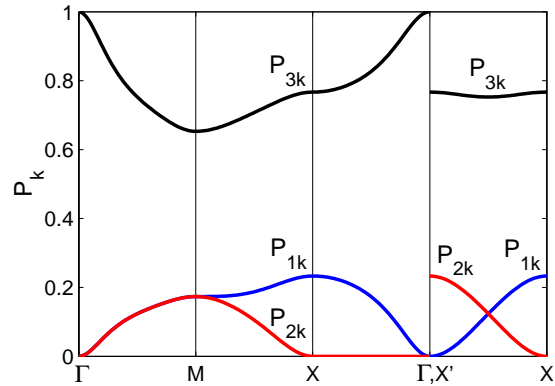}
\caption{\label{Falf} Partial contributions of the basis states to
the one-hole state corresponding to the lower branch of the
spectrum in Fig. \ref{Fsp}b. The model parameters are as in Fig.
\ref{Fsp}. $\Gamma=(0,0)$, $M=(\pi,\pi)$, $X=(0,\pi)$, and
$X'=(\pi,0)$.}
\end{center}
\end{figure}

Let us consider the structure of the one-hole state illustrated by
the lower branch of the spectrum in Fig. 1b. Weight contributions
$P_{1k}$ and $P_{2k}$ of bare hole states
$A^+_{1k\sigma}|G\rangle$ and $A^+_{2k\sigma}|G\rangle$ are
determined as $P_{1k}=|\alpha_{1k}|^2$ and
$P_{2k}=|\alpha_{2k}|^2$. The weight contribution of the spin
polaron basis state is $P_{3k}=K_{33}|\alpha_{3k}|^2 $.

Figure \ref{Falf} shows the partial contributions for the
quasimomenta lying on the four directions of the Brillouin zone.
It can be seen that $P_{3k}$ (upper curves) multiply exceeds
$P_{1k}$ and $P_{2k}$, which proves the spin polaron nature of the
one-hole state corresponding to the lower split branch of the
spectrum.

The dependencies presented in Fig. \ref{Fsp} were calculated with
the use of the dispersion equation
\begin{eqnarray}\label{det3}
{\det}_k(\omega) =\left|D(k)-\omega K(k)\right|=0,
\end{eqnarray}
obtained from the condition of nontriviality of the solutions of
system (\ref{sysEqVar}). Developing the determinant, we arrive at:
\begin{eqnarray}\label{detN}
{\det}_k(\omega) =\left(\omega-\varepsilon_p\right)^3-
Q_k\left(\omega-\varepsilon_p\right)^2+
B_k\left(\omega-\varepsilon_p\right)+R_k, \nonumber
\end{eqnarray}
where we made the following designations:
\begin{eqnarray}\label{definit3}
&& Q_k=2\tau_-(1+\gamma_{1k})+\Lambda_k,\nonumber\\
&&B_k= \left( 2\tau_-\Lambda_k
-16\tau_+^2K_{33}\right)(1+\gamma_{1k})+ 4t\left(\tau_--
t\right)\chi_k,\nonumber\\
&&R_k=4t\chi_k\left[ 8\tau_+^2K_{33}-
\Lambda_k (\tau_- -t)\right],\nonumber\\
&&\Lambda_k=\frac{D_{33}}{K_{33}}-\varepsilon_p,~~\chi_k=
1+2\gamma_{1k}+\gamma_{2k}.
\end{eqnarray}

For the set of parameters used, it is easy to obtain the
approximate solutions of the dispersion equation that describe the
one-hole state spectrum with high accuracy. In this case, the spin
polaron spectrum is determined as:
\begin{eqnarray}
\epsilon_{sp}(k)=
\varepsilon_p+x_k,~x_k=\frac{Q_k}{2}-\sqrt{\frac{Q_k^2}{4}-B_{k}-\frac{R_{k}}{x_{k0}}},~
\end{eqnarray}
where $x_{k0}=Q_k/2-\sqrt{Q_k^2/4-B_{k}}$. For the two upper
branches, obtain:
\begin{eqnarray}
\epsilon_{1k}=\varepsilon_p+\frac{Q_k-x_k}{2}-\nu_k,~
\epsilon_{2k}=\varepsilon_p+\frac{Q_k-x_k}{2}+\nu_k,
\nonumber\\
\nu_k=\sqrt{(Q_k-x_k)^2/4+R_k/x_k}.~~~~~~~~~~~~~~~~~
\end{eqnarray}
\bigskip

4. \textbf{Equations of motion for the spin polaron Green's
functions in the superconducting phase}

Developing of the theory of the superconducting state in a system
of oxygen holes strongly coupled with the subsystem of the
localized spin moments of copper ions suggests expanding of the
set of basis operators that would allow introducing anomalous
averages. Taking into account the results reported in Section 3,
we can easily see that the formulated problem can be solved by
adding operators $A^+_{1,-k,\bar\sigma}$, $A^+_{2,-k,\bar\sigma}$
and $A^+_{3,-k,\bar\sigma}$ to operators $A_{1k\sigma}$,
$A_{2k\sigma}$ and $A_{3k\sigma}$ used previously.

To obtain self-consistency equations in the superconducting phase,
we use the introduced basis of six operators and apply the
Zwanzig--Mori technique
\cite{Plakida_book_10,Plakida03,Vladimirov07,Zwanzig1961,Mori1965}.

Let us introduce the retarded two-time temperature Green's
functions (GPs) ($i,j=1,\dots,6$):
\begin{eqnarray}  \label{GF}
G_{ij}(k,t)=\langle\langle
A_{ik\sigma}(t)|A^+_{jk\sigma}(0)\rangle\rangle
=\nonumber\\
=-i\theta(t)\langle [ A_{ik\sigma}(t),A^+_{jk\sigma}(0)]\rangle.
\end{eqnarray}

The system of $6\times 6$ equations of motion for these GPs is:
\begin{eqnarray}  \label{EqMGF}
\omega\langle\langle
A_{ik\sigma}|A^+_{jk\sigma}\rangle\rangle_\omega=
K_{ij}(k)+\langle\langle [A_{ik\sigma},\hat{\mathscr
H}~]|A^+_{jk\sigma}\rangle\rangle_\omega,
\end{eqnarray}
where the free term $K_{ij}(k)=\langle \{
A_{ik\sigma},A^+_{jk\sigma}\}\rangle$ is determined analogously to
expression (\ref{Kij}), but with averaging over the ensemble of
spin polaron quasiparticles. The aforesaid is valid also to the
matrix elements $D_{ij}(k)=\langle\{[A_{ik\sigma},\hat{\mathscr
H}~],A^+_{jk\sigma}\}\rangle$. At the low doping level, the values
of matrix elements $K_{ij}(k)$ and $D_{ij}(k)$ for the indices
$i,j=1,\dots,3$ are determined by expressions (\ref{K13}) and
(\ref{D13}).

In the projection technique, the Green's functions obtained by the
commutation $[A_{ik\sigma},\hat{\mathscr H}]$ are written as the
linear superposition of basis GPs (\ref{GF}), i.e.:
\begin{eqnarray}\label{MoriAp}
\langle\langle [A_{ik\sigma},\hat{\mathscr
H}~]|A^+_{jk\sigma}\rangle\rangle_\omega = \sum_l
L_{il}(k)\langle\langle
A_{lk\sigma}|A^+_{jk\sigma}\rangle\rangle_\omega,
\end{eqnarray}
where $L(k)=D(k)K^{-1}(k)$. The obtained system of equations for
the GPs $\langle\langle
A_{ik\sigma}|A^+_{jk\sigma}\rangle\rangle_\omega$ is closed and
can be presented, for the sake of briefness, in the matrix form:
\begin{eqnarray}
\label{sys} \left(\omega\cdot \hat
I-D(k)K^{-1}(k)\right)G(k,\omega)=K(k),
\end{eqnarray}
where $\hat I$ is the unit matrix. Then, the energy spectrum of
quasiparticles $E_{jk}$ in the superconducting phase is determined
by poles of GP $G$ and can be obtained from the six-order
dispersion equation:
\begin{eqnarray}\label{det0}
\det\left|~\omega\cdot \hat I -D(k)K^{-1}(k)\right|=0.
\end{eqnarray}

Matrix elements $K_{ij}(k)$ and $D_{ij}(k)$ with the indices
$i,j=1,\dots,3$ are known (see (\ref{K13}) and (\ref{D13})).
Matrix ${K}(k)$ is still diagonal, and besides
$K_{j+3,j+3}(k)=K_{jj}(k)$ ($j=1,\dots,3$). Matrix ${D}(k)$ is
convenient to be written in the block representation. The left
upper block $3 \times 3$ in size is composed only from the normal
averages $D_{ij}(k)$ ($i,j=1,2,3$). The right lower block is
formed of $D_{ij}(k)$ with $i,j=4,5,6$. Furthermore
$D_{i+3,j+3}(k)=-D_{ij}(k)$ for all $i,j=1,2,3$.

Matrix elements in the right upper block of ${D}(k)$ are formed
due to anomalous pairings. In our case, only element $D_{36}(k)$
in the block is nonzero. It follows from the hermiticity condition
that in the left lower block of ${D}(k)$, only element $D_{63}(k)$
coinciding with $D_{36}(k)$ is nonzero.

In view of the aforesaid, dispersion equation (\ref{det0}) can be
written as:
\begin{eqnarray}\label{detSC}
{\det}_k(\omega){\det}_k(-\omega)+
\varphi_k(\omega)\varphi_k(-\omega)\left|\frac{D_{36}(k)}{K_{33}(k)}
\right|^2=0,
\end{eqnarray}
where
\begin{eqnarray}
&&\varphi_k(\omega)=(\omega+\varepsilon_p)^2+2\tau_-(1+\gamma_1(k))
(\omega+\varepsilon_p)+
\nonumber\\
&&+4t(\tau_--t)\chi(k).
\end{eqnarray}
In the normal phase, $D_{36}(k)=0$ and Eq. (\ref{detSC}) reduces
to Eq. (\ref{det3}).
\bigskip

5. \textbf{Self-consistency equation for the superconducting order
parameter}

Anomalous average $D_{36}(k)$ is expressed via a sum of a large
number of terms that can yield the solutions of the integral
self-consistency equation with different types of symmetry of the
superconducting order parameter (SOP). In particular, the terms
proportional to parameter $\tau_+$ yield $s$-wave pairing. In view
of the experimental data, we limit the consideration to the
$d$-type SOP. In this case, we may use the truncated expression:
\begin{eqnarray}\label{D36}
D_{36}(k)=I_1\sum_{\delta}e^{ik2\delta}\left[-\langle
A^+_{6f\sigma}A_{3,f+2\delta,\sigma}\rangle+\right.
\nonumber\\
+\left. \frac{C_1}{4}\sum_{\delta'\delta_1} \langle
c_{f+2\delta+\delta_1,\bar\sigma}~ c_{f+\delta',\sigma}\rangle
\right].
\end{eqnarray}
We obtained this expression by applying the decoupling procedure
to the averages with the products of operators that cannot be
reduced to the basis operators. This causes the occurrence of
magnetic correlator $C_1$ in front of the second term expressed
via a sum of the averages of the introduced basis operators.

From expression (24), obtain
\begin{eqnarray}
D_{36}(k)=\Delta_0(\cos k_x - \cos k_y).
\end{eqnarray}
The amplitude of SOP $\Delta_0$ is determined from the equation:
\begin{eqnarray}\label{EqD}
1=\frac{I_1}{N}\sum_k\frac{(\cos k_x-\cos
k_y)^2}{2E_{k}\left(E_{k}^2-\epsilon_{1k}^2\right)\left(E_{k}^2-\epsilon_{2k}^2\right)}
\tanh\left(\frac{E_{k}}{2T} \right) \nonumber\\
\times\left[ \varphi_k(E_{k})\varphi_k(-E_{k})
-16C_1\tau_+^2\Psi_k(E_{k})\Psi_k(-E_{k})\right],~
\end{eqnarray}
where
\begin{eqnarray}
\Psi_k(\omega)=(\omega-\varepsilon_p)(1+\gamma_1(k))-2t\chi(k),
\end{eqnarray}
and the Fermi excitation spectrum $E_k$ in the superconducting
phase is:
\begin{eqnarray}
E_k=\sqrt{\epsilon^2_{sp}(k)+D^2_{36}(k)}.
\end{eqnarray}
This formula demonstrates that in the superconducting phase the
Fermi excitation spectrum is based on the spectrum of the spin
polaron states. Therefore, we may state that the investigated
Cooper instability describes the instability of the spin polaron
ensemble relative to the Cooper pairing.
\bigskip

6. \textbf{Effect of doping on the Cooper instability of the spin
polaron ensemble}

The concentration dependence of the critical temperature of the
transition to the superconducting phase with the d-type
order-parameter symmetry was calculated using expression
(\ref{EqD}) with regard to the equation for the chemical potential
\begin{eqnarray}\label{Eqmu}
\frac{x}{4}=\frac{1}{N}\sum_k\frac{J_k(E_k)f(E_k/T)-J_k(-E_k)f(-E_k/T)}
{2E_{k}\left(E_{k}^2-\epsilon_{2k}^2\right)\left(E_{k}^2-\epsilon_{3k}^2\right)},
\end{eqnarray}
where the function $J_k(\omega)$ is expressed as
\begin{eqnarray}\label{Jdef}
J_k(\omega)=\left(-\omega+\varepsilon_p+\tau_-(1+\gamma_{1k})\right)\varphi_k(\omega)
\frac{|D_{36}(k)|^2}{K_{33}(k)}-
\nonumber\\
-\left[\left(\omega-\varepsilon_p-\tau_-(1+\gamma_{1k})\right)\left(\omega-D_{33}(k)/K_{33}(k)
 \right)-   \right.
 \nonumber\\
 -8\tau^2_+K_{33}(k)(1+\gamma_{1k}){\Large\left. \right]}{\det}_k(-\omega),~~~~~
\end{eqnarray}
and $f(z)=1/(\exp(z)+1)$ is the Fermi--Dirac function. In the
numerical calculations, it was taken into account that $K_{ij}$
and $D_{ij}$ depend on spin correlation functions $C_j$
(\ref{spcor}) for the first three coordination spheres $j=1,2,3$.
These correlation functions, as well as the gap in the magnetic
excitation spectrum in the vicinity of point $\mathbf{Q}$ of the
Brillouin zone were jointly determined within the spherical
symmetric self-consistent approach for a frustrated ferromagnet
\cite{BarMikhShvar2011}. Determination of the spin correlation
functions with the use of the model considered here was described
in detail in \cite{DVB_2013}.

The change in hole concentration $x$ was taken into account via
both modification of spin correlators $C_j$ and shifting of
chemical potential $\mu$.

\begin{figure}
\begin{center}
\includegraphics[width=240pt, height=220pt, angle=0, keepaspectratio]{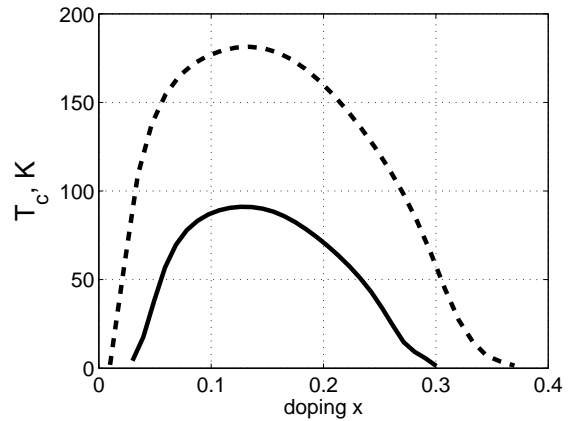}
\caption{\label{FTc} Concentration dependence of the temperature
of the transition to the superconducting state with the d-type
symmetry of the SOP. Dashed line corresponds to $I=0.34\tau$ and
solid line, to $I=0.2\tau$. The rest parameters are as in Fig.
\ref{Fsp}.}
\end{center}
\end{figure}

The calculated data are presented in Fig. \ref{FTc}.
It can be seen that the region of implementation of the
superconducting phase with the d-type order-parameter symmetry
depends on the exchange interaction value. This results from the
fact that the Cooper instability manifests itself in the ensemble
of quasiparticles that are spin polarons. Thus, the exchange
interaction between spins of copper ions induces attraction
between these complex Fermi quasiparticles. Note that for the
characteristic values of the exchange integral, the obtained phase
diagram is in good agreement with the available experimental data
on both hole concentration $x$ and critical temperature $T_c$.

\begin{figure}
\begin{center}
\includegraphics[width=240pt, height=220pt, angle=0, keepaspectratio]{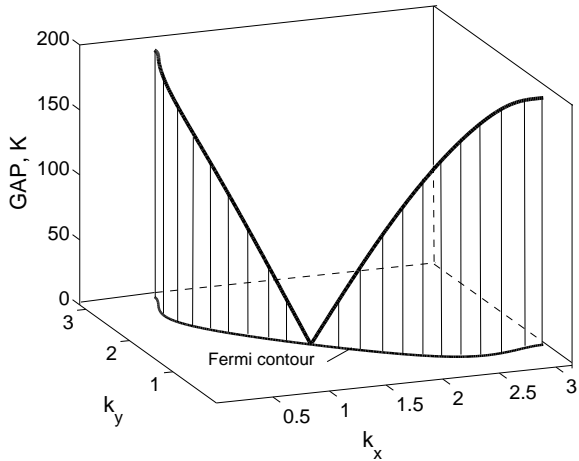}
\caption{\label{FD} Superconducting gap vs quasimomentum on the
Fermi contour. The Fermi contour is indicated by the solid line in
the horizontal plane. The calculation was made for $x=0.125$,
$I=0.2\tau$ and $T=0$. The rest parameters are as in Fig.
\ref{Fsp}.}
\end{center}
\end{figure}

Figure \ref{FD} shows the gap variation in the spectrum of
elementary excitations of spin polaron quasiparticles on the Fermi
contour in the superconducting phase. It can be seen that the
dependence of the gap on the quasimomentum in the first Brillouin
zone is characterized by the d-type symmetry.

\begin{figure}[h]
\begin{center}
\includegraphics[width=220pt, height=220pt, angle=0, keepaspectratio]{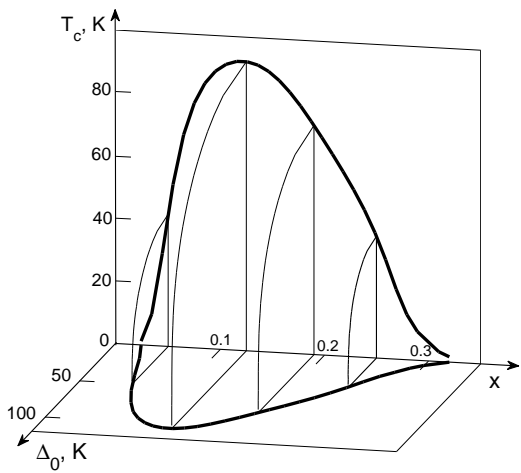}
\caption{\label{FTDx} Variation in the amplitude of the
superconducting order parameter and temperature of the transition
to the superconducting phase upon doping. The model parameters are
as in Fig. \ref{Fsp}.}
\end{center}
\end{figure}

Figure \ref{FTDx} presents concentration dependences of order
parameter $|D_{36}(k)|$ and critical temperature $T_c$. As follows
from this figure, the amplitude of the order parameter turns to
zero at approaching $T_c$ by the second-order phase transition.
\bigskip

8. \textbf{Conclusions}

In this study, we obtained the following results.

(i) Using the spin-fermion model that takes into account the
strong correlation between charge and spin degrees of freedom and
the real structure of the CuO$_2$-plane with two oxygen ions per
unit cell, it was demonstrated for the first time that with a
decrease in temperature the ensemble of spin polaron
quasiparticles passes to the superconducting state with the d-type
symmetry of the order parameter.

(ii) The mechanism ensuring Cooper pairing of spin polarons is the
exchange interaction, which transforms to effective attraction
between spin polarons as a result of the strong spin-charge
coupling. In this regard it should be emphasized the important
role of the spin-flip processes in the formation of the Cooper
instability.

(iii) Despite the complexity of the initial self-consistency
equations, simple analytical expressions were obtained for both
the spin polaron spectrum in the normal phase and the Fermi
excitation spectrum in the superconducting phase.

(iv) The $T-x$-diagram obtained using the spin polaron concept is
in good agreement with the experimental data for copper oxides.
\bigskip

\textbf{Acknowledgments}

This study was supported by the Russian Foundation for Basic
Research, projects nos. 13-02-00523 and 13-02-00909 and the
Dynasty Foundation.

\bigskip

\end{document}